\documentclass[aps,showpacs,preprintnumbers,nofootinbib]{revtex4-1}  \usepackage{slashed,color,amsmath,amssymb,mathrsfs}
\usepackage{graphicx} 
  \allowdisplaybreaks

\begin{document} \preprint{} \title{Higgsphobic and fermiophobic Z' as a single dark matter candidate}

\author{Nan Chen$^1$} \email[]{chen-n08@mails.tsinghua.edu.cn} \author{Ying Zhang$^2$} \email[]{hepzhy@mail.xjtu.edu.cn} \author{Qing
Wang$^{1,3,4}$} \email[Corresponding author ]{wangq@mail.tsinghua.edu.cn} \affiliation{$^1$Department of Physics, Tsinghua University, Beijing
100084, P.R. China\\ $^2$School of Science, Xi'an Jiaotong University, Xi'an 710049, P.R. China\\ $^3$Center for High Energy Physics, Tsinghua
University, Beijing 100084, P.R. China\\ $^4$Collaborative Innovation Center of Quantum Matter, Beijing 100084, P.R. China} \author{Giacomo
Cacciapaglia$^{5}$} \email[Corresponding author ]{g.cacciapaglia@ipnl.in2p3.fr} \author{Aldo Deandrea$^{5,6}$} \email[]{deandrea@ipnl.in2p3.fr}
\author{Luca Panizzi$^7$} \email[]{l.panizzi@soton.ac.uk} \affiliation{$^5$Universit$\acute{e}$ de Lyon, F-69622 Lyon, France;
Universit$\acute{e}$ Lyon 1, Villeurbanne\\ CNRS/IN2P3, UMR5822, Institut de Physique Nucl$\acute{e}$aire de Lyon F-69622 Villeurbanne Cedex,
France\\ $^6$ Institut Universitaire de France, 103 boulevard Saint-Michel, 75005 Paris, France\\ $^7$School of Physics and Astronomy,
University of Southampton, Highfield, Southampton SO17 1BJ, UK}
 \begin{abstract}
A spin-1 Z' particle as a single dark matter candidate is investigated by assuming that it does not directly couple to the Higgs boson and
standard model fermions and does not mix with the photon and Z boson. The remaining dominant vertices are quartic $Z'Z'ZZ$ and $Z'Z'W^+W^-$,
which can induce effective $Z'Z'q\bar{q}$ couplings through standard-model gauge-boson loops. We discuss constraints from the cosmological
thermal relic density, and direct and indirect-detection experiments, and find that a dark Z' can only exist above the W boson mass threshold,
and the effective quartic coupling of $Z'Z'VV$ is bounded in the region of $10^{-3}\sim 10^{-2}$. \end{abstract} 
\maketitle
\section{Introduction}

After the discovery of the 125~GeV Higgs boson, the standard model (SM) of particle physics has become a complete theory; within the SM, the
remaining task is the precision measurements of various Higgs properties, in particular its couplings, and to further narrow down the possible
parameter space of new physics. Although the hierarchy and meta-stable vacuum problems remain for the SM Higgs theoretically, for new physics
beyond the SM the absence of new physics signals at the LHC to date implies that extensions of the SM still only need to rely on the traditional
particle physics facts of non-zero neutrino masses and baryon asymmetry. Given these circumstances, the presence of dark matter (DM) in our
Universe becomes an even more important leading empirical evidence for the existence of new physics, because no SM particle can account for DM.
Cosmology and astrophysics tell us that almost 85$\%$ of matter in our universe is dark, i.e., neutral, non-luminous and non-baryonic. The fact
that the abundance of DM is comparable to that of ordinary visible matter seems to imply that DM may have the same or similar origins and
properties as ordinary matter. If we accept the conclusion of quantum field theory (QFT) that all matter should be made of particles, then an
unambiguous, non-gravitational signal of DM must appear in particle physics experiments. This has driven the particle physics community to try
harder to unravel DM's still enigmatic properties.

Because details of the particle properties of DM are lacking, the best investigative strategy for theorists is to try to cover as much ground as
possible. Considering that QFT classifies particles according to their spin (even or odd half-integers), elementary particles discovered so far
all have low spins. Most DM candidates discussed so far in the literature have been assumed to be spin-0 scalars
\cite{ScalarDM1,*ScalarDM2,*ScalarDM3,*ScalarDM4,*ScalarDM5,*ScalarDM6,*ScalarDM7} or spin-1/2 spinors \cite{SpinorDM3,*SpinorDM4}. Whereas a
scalar DM has relatively simple structure and provides possible intimate interplay with the 125~GeV Higgs, a spinor DM extends the traditional
observation that matter is composed of spin-1/2 particles. The heavy sterile neutrinos \cite{SpinorDM1} and the lightest neutralino
\cite{SpinorDM2} in supersymmetric models are DM candidates belonging to this type. Apart from scalar and spinor DM, the next level of
higher-spin candidate particles comprise spin-1 vectors. If we limit ourselves to the simplest vector particle scenario in particle physics, a
single extra neutral vector particle, usually denoted by Z', is sufficient. We shall discuss this possibility in the present paper. A higher
spin case, spin-3/2 DM, has been discussed in Ref.\cite{spin3over2}.

A vector particle Z' can be viewed as a gauge boson that mediates an extra $U(1)$ gauge force beyond the conventional SM strong $SU(3)_c$ force
and electroweak $SU(2)_L\otimes U(1)_Y$ forces. For as yet unknown reasons, this additional $U(1)$ gauge symmetry is spontaneously broken, thus
yielding a massive Z'. SM plus Z' is a minimal and well motivated generalisation of the SM; many new-physics models have such a $Z'$ boson (for
details see review Refs. \cite{LangackerRMP2008,*LangackerPRL2008,*Han,*Zwirner}) as a necessary constituent and remnant for new-physics
interactions. Before July 4, 2012, the Higgs was the superstar of particle physics searches and a Z' only played a supporting role. With the
discovery of the 125~GeV Higgs, a Z' now becomes one of the hot new-physics candidate particles and the LHC is actively searching for it in
various channels, with the model-dependent lower mass bound already reaching the TeV-energy region depending on the final state it is assumed to
decay into. Now, if we further take the Z' as a DM candidate thus changing the Z' from visible to invisible, the interactions between the
invisible Z' and SM particles will be strongly reduced, and the corresponding search strategies (such as direct detection, indirect detection,
and collider experiment) will change with respect to those for visible Z'. Various constraints must therefore be re-examined.

In the literature, the invisible Z' has been intensively discussed as a messenger between the visible sector (which contains the SM particles)
and a hidden sector (to which DM belongs) \cite{DMZ1,*DMZ2,*DMZ3,*DMZ4,*DMZ5,*DMZ6,*DMZ7,*DMZ8,*DMZ9,*DMZ10,*DMZ11,*DMZ12}: in such a scenario,
the SM particles can be either charged under the additional gauge symmetry or not. In the event that SM particles are neutral with respect to
the extra $U(1)$ symmetry, the interaction occurs via effective operators connecting directly Z' to the SM sector. The simplest case is the
kinetic mixing terms between the SM hypercharge field strength and the new Abelian field strength \cite{mixing}. The underlying reason in
adopting Z' as a portal to the hidden sector stems from the traditional mediating role of gauge bosons. In this type of DM models, there are too
many unknowns concerning the hidden sector, a situation that is not helpful in DM searches. In this paper, we consider an alternative simple
approach by ignoring the conventional messenger role of Z', and instead treat it as pure matter. This approach is similar to the minimal darkon
model \cite{ScalarDM1,*ScalarDM2,*ScalarDM3,*ScalarDM4,*ScalarDM5,*ScalarDM6} where SM is minimally expanded with the addition of a dark scalar
(SM+D), except now we replace the scalar darkon D with a single vector DM candidate Z'. The change from the traditional Z' portal model to our
present single dark Z' approach is similar to that from the Higgs portal model (where a scalar is taken as a messenger between the visible and
hidden sectors) \cite{HiggsPortal} to the darkon model. After the reduction, because of the unique choice of DM candidate, we can ignore the
uncertainties arising from the arbitrary hidden sector in the traditional Z' or Higgs portal models. The difference in the present approach with
respect to scalar DM is that our dark Z' is a vector particle, which behaves not like a scalar or Higgs boson, but very much like a Z boson of
SM and will have relatively complex interaction structure owing to its polarisation. A spin-1 dark matter candidate appears in models with one
extra dimensions~\cite{KK5} and has been widely studied in this context~ \cite{*KKDM1,*KKDM2}. Note that this is not a generic prediction of
extra dimensions, as in higher then 5 dimensions the candidate is a scalar~\cite{*KK6a,*KK6b,*KK6c}, and a scalar is again found in models of
pseudo-Goldstone Higgs in warped space~\cite{frigerio} and technicolor~\cite{sannino}.

This paper is organised as follows. In Section II, in terms of the model-independent extended electroweak chiral Lagrangian and the six
assumptions needed to keep Z' dark, we determine the necessary operators that couple our dark Z' to SM particles. In Section III, we calculate
the relic density produced from our single dark Z', derive a constraint on the effective coupling of the dark Z' pair to $W$ or $Z$ pairs.
Section IV looks at the direct-detection constraint, where we compute the SM gauge-boson-loop-induced $Z'Z'\bar{q}q$ vertex and discuss direct
detection. Section V examines indirect-detection constraints and includes discussions of the Pamela, AMS02, and FermiLAT experiments. In Section
VI, we discuss the combined results and some other possible DM related issues. Section VII presents a summary. Some necessary results for
Section II are to be found in Appendix A.

\section{$SU(2)_L\otimes U(1)_Y\otimes U(1)$ theory and Z' as a DM candidate}

To make our investigation general, we start from a model-independent effective extended electroweak chiral Lagrangian (EEWCL) proposed in
Ref.\cite{CL}, \begin{eqnarray} \mathcal{L}_{\mathrm{EEWCL}}=\mathcal{L}_2+\mathcal{L}_4+\cdots, \end{eqnarray} where $\mathcal{L}_{2n}$ for
$n=1,2,3,\ldots$ is the $p^{2n}$-order of EEWCL with Z' and all SM gauge fields, plus four necessary Goldstone bosons described by a two-by-two
unitary matrix field $\hat{U}$. The symmetry of the Lagrangian is $SU(2)_L\otimes U(1)_Y\otimes U(1)$, which will be spontaneously broken to
$U(1)_{\mathrm{em}}$. The SM Higgs field and the fermion fields are not included in the above Lagrangian: in fact, couplings to the Higgs are
not required by the symmetry of the model and, if present, would reproduce a model of Higgs portal DM. This implies that we ignore the possible
direct (or tree-order) couplings between Z' and Higgs (a dark Z' with direct coupling to the Higgs has been discussed in Ref.\cite{ZprimeHiggs})
or between Z' and SM fermions (a dark Z' coupling directly to SM fermions has been discussed in Ref.\cite{Yu:2011by}). These are the first two
assumptions we adopt for our dark Z'. These higgsphobic and fermiophobic dark Z' assumptions simplify our theory significantly, and we take it
as the first step in our investigation. Although at tree level we can ignore the explicit Higgs and fermion couplings, loops can still induce
effective couplings. We shall carefully discuss these loops in Section IV.

The $p^2$-order Lagrangian $\mathcal{L}_2$ is \cite{CL} \begin{eqnarray} \mathcal{L}_2 &=& -\frac{1}{4}f^2\mathrm{tr}[\hat{V}_\mu\hat{V}^\mu]
+\frac{1}{4}\beta_1f^2\mathrm{tr}[T\hat{V}_\mu]\mathrm{tr}[T\hat{V}^\mu] +\frac{1}{4}\beta_2f^2\mathrm{tr}[\hat{V}_\mu]\mathrm{tr}[T\hat{V}^\mu]
+\frac{1}{4}\beta_3f^2\mathrm{tr}[\hat{V}_\mu]\mathrm{tr}[\hat{V}^\mu], \end{eqnarray} where $T\equiv\hat{U}\tau_3\hat{U}^\dag$,
$\hat{V}_\mu\equiv(D_\mu\hat{U})\hat{U}^\dag$, and $\tau_3$ is the Pauli matrix. The covariant derivative is \begin{eqnarray}
D_\mu\hat{U}=\partial_\mu\hat{U}+igW_\mu\hat{U}-i\hat{U}\frac{\tau_3}{2}g'B_\mu -i\hat{U}(\tilde{g}'B_\mu+g^{\prime\prime}X_\mu)I,\label{DUdef}
\end{eqnarray} where $W_\mu\equiv\frac{\tau_i}{2}W^i_\mu$, $B_{\mu}$, and $X_\mu$ are the $SU(2)_L$, $U(1)_Y$, and $U(1)$ gauge fields,
respectively, and $g,g',g^{\prime\prime}$ are the corresponding coupling constants; $\tilde{g}'$ is a special Stueckelberg coupling. $W_\mu$,
$B_{\mu}$, and $X_\mu$ are gauge eigenstates, and the Z' discussed in this paper is the physical state after diagonalization. In
$\mathcal{L}_2$, $\beta_1$ and $\beta_2$ are mass mixing parameters.

The $p^4$-order Lagrangian $\mathcal{L}_4$ is composed of three terms\cite{CL} \begin{eqnarray}
\mathcal{L}_4&=&\mathcal{L}_K+\mathcal{L}_B+\mathcal{L}_A\label{Ldef}, \end{eqnarray} for which the kinetic term $\mathcal{L}_K$ is
\begin{eqnarray} \mathcal{L}_K&=&
    -\frac{1}{4}B_{\mu\nu}B^{\mu\nu}
    -\frac{1}{2}\mathrm{tr}[W_{\mu\nu}W^{\mu\nu}]
    -\frac{1}{4}X_{\mu\nu}X^{\mu\nu}.
\end{eqnarray} The normal term $\mathcal{L}_B$ is \begin{eqnarray} \mathcal{L}_B &=&\frac{1}{2}\alpha_1gg'B_{\mu\nu}\mathrm{tr}[TW^{\mu\nu}]
+\frac{i}{2}\alpha_2g'B_{\mu\nu}\mathrm{tr}[T[\hat{V}^\mu,\hat{V}^\nu]] +i\alpha_3g\mathrm{tr}[W^{\mu\nu}[\hat{V}^\mu,\hat{V}^\nu]]\nonumber\\
&&+\alpha_4\mathrm{tr}[\hat{V}_\mu\hat{V}_\nu]\mathrm{tr}[\hat{V}^\mu\hat{V}^\nu]
+\alpha_5\mathrm{tr}[\hat{V}_\mu\hat{V}^\mu]\mathrm{tr}[\hat{V}^\nu\hat{V}_\nu]
+\alpha_6\mathrm{tr}[\hat{V}_\mu\hat{V}_\nu]\mathrm{tr}[T\hat{V}^\mu]\mathrm{tr}[T\hat{V}^\nu]\nonumber\\
&&+\alpha_7\mathrm{tr}[\hat{V}_\mu\hat{V}^\mu]\mathrm{tr}[T\hat{V}_\nu]\mathrm{tr}[T\hat{V}^\nu]
+\frac{1}{4}\alpha_8g^2\mathrm{tr}[TW_{\mu\nu}]\mathrm{tr}[TW^{\mu\nu}] +\frac{i}{2}\alpha_9
g\mathrm{tr}[TW^{\mu\nu}]\mathrm{tr}[T[\hat{V}_\mu,\hat{V}_\nu]]\nonumber\\
&&+\frac{1}{2}\alpha_{10}\mathrm{tr}[T\hat{V}^\mu]\mathrm{tr}[T\hat{V}^\nu]\mathrm{tr}[T\hat{V}_\mu]\mathrm{tr}[T\hat{V}_\nu]
+\alpha_{11}g\epsilon^{\mu\nu\rho\lambda}\mathrm{tr}[T\hat{V}_\mu]\mathrm{tr}[\hat{V}_\nu W_{\rho\lambda}]\nonumber\\
&&+\alpha_{12}g\mathrm{tr}[T\hat{V}^\mu]\mathrm{tr}[\hat{V}^\nu W_{\mu\nu}]
+\alpha_{13}gg'\epsilon^{\mu\nu\rho\lambda}B_{\mu\nu}\mathrm{tr}[TW_{\rho\lambda}]
+\alpha_{14}g^2\epsilon^{\mu\nu\rho\lambda}\mathrm{tr}[TW_{\mu\nu}]\mathrm{tr}[TW_{\rho\lambda}]\nonumber\\
&&+\alpha_{15}\mathrm{tr}[\hat{V}_\mu]\mathrm{tr}[T\hat{V}^\mu]\mathrm{tr}[T\hat{V}_\nu]\mathrm{tr}[T\hat{V}^\nu]
+\alpha_{16}\mathrm{tr}[\hat{V}_\mu]\mathrm{tr}[T\hat{V}^\mu]\mathrm{tr}[\hat{V}_\nu\hat{V}^\nu]
+\alpha_{17}\mathrm{tr}[\hat{V}_\mu]\mathrm{tr}[T\hat{V}_\nu]\mathrm{tr}[\hat{V}^\mu\hat{V}^\nu]\nonumber\\
&&+\alpha_{18}\mathrm{tr}[\hat{V}_\mu]\mathrm{tr}[\hat{V}_\nu]\mathrm{tr}[T\hat{V}^\mu]\mathrm{tr}[T\hat{V}^\nu]
+\alpha_{19}\mathrm{tr}[\hat{V}_\mu]\mathrm{tr}[\hat{V}_\nu]\mathrm{tr}[\hat{V}^\mu\hat{V}^\nu]
+\alpha_{20}\mathrm{tr}[\hat{V}_\mu]\mathrm{tr}[\hat{V}^\mu]\mathrm{tr}[T\hat{V}_\nu]\mathrm{tr}[T\hat{V}^\nu]\nonumber\\
&&+\alpha_{21}\mathrm{tr}[\hat{V}_\mu]\mathrm{tr}[\hat{V}^\mu]\mathrm{tr}[\hat{V}_\nu\hat{V}^\nu]
+\alpha_{22}\mathrm{tr}[\hat{V}_\mu]\mathrm{tr}[\hat{V}^\mu]\mathrm{tr}[\hat{V}_\nu]\mathrm{tr}[T\hat{V}^\nu]
+\alpha_{23}\mathrm{tr}[\hat{V}_\mu]\mathrm{tr}[\hat{V}_\nu]\mathrm{tr}[\hat{V}^\mu]\mathrm{tr}[\hat{V}^\nu]\nonumber\\
&&+gg^{\prime\prime}\alpha_{24}X_{\mu\nu}\mathrm{tr}[TW^{\mu\nu}] +g'g^{\prime\prime}\alpha_{25}B_{\mu\nu}X^{\mu\nu}
+\alpha_{26}\epsilon^{\mu\nu\rho\lambda}\mathrm{tr}[\hat{V}_\mu]\mathrm{tr}[T\hat{V}_\nu]\mathrm{tr}[T[\hat{V}_\rho,\hat{V}_\lambda]]\nonumber\\
&&+ig'\alpha_{27}\epsilon^{\mu\nu\rho\lambda}\mathrm{tr}[\hat{V}_\mu]\mathrm{tr}[T\hat{V}_\nu]B_{\rho\lambda}
+ig\alpha_{28}\epsilon^{\mu\nu\rho\lambda}\mathrm{tr}[\hat{V}_\mu]\mathrm{tr}[T\hat{V}_\nu]\mathrm{tr}[TW_{\rho\lambda}]\nonumber\\
&&+g\alpha_{29}\epsilon^{\mu\nu\rho\lambda}\mathrm{tr}[\hat{V}_\mu]\mathrm{tr}[\hat{V}_\nu
W_{\rho\lambda}]+ig^{\prime\prime}\alpha_{30}\epsilon^{\mu\nu\rho\lambda}X_{\mu\nu}\mathrm{tr}[T[\hat{V}_\rho,\hat{V}_\lambda]]
+ig^{\prime\prime}\alpha_{31}X_{\mu\nu}\mathrm{tr}[T[\hat{V}^\mu,\hat{V}^\nu]]\nonumber\\
&&+g^{\prime\prime}\alpha_{32}\epsilon^{\mu\nu\rho\lambda}\mathrm{tr}[\hat{V}_\mu]\mathrm{tr}[T\hat{V}_\nu]X_{\rho\lambda}
+\alpha_{33}\mathrm{tr}[\hat{V}_\mu]\mathrm{tr}[T\hat{V}_\nu]\mathrm{tr}[T[\hat{V}^\mu,\hat{V}^\nu]]
+g'g^{\prime\prime}\alpha_{34}\epsilon^{\mu\nu\rho\lambda}B_{\mu\nu}X_{\rho\lambda}\nonumber\\
&&+gg^{\prime\prime}\alpha_{35}\epsilon^{\mu\nu\rho\lambda}X_{\mu\nu}\mathrm{tr}[TW_{\rho\lambda}]
+ig'\alpha_{36}\mathrm{tr}[\hat{V}_\mu]\mathrm{tr}[T\hat{V}_\nu]B^{\mu\nu}
+ig\alpha_{37}\mathrm{tr}[\hat{V}_\mu]\mathrm{tr}[T\hat{V}_\nu]\mathrm{tr}[TW^{\mu\nu}]\nonumber\\
&&+g\alpha_{38}\mathrm{tr}[\hat{V}^\mu]\mathrm{tr}[\hat{V}^\nu W_{\mu\nu}]
+g^{\prime\prime}\alpha_{39}\mathrm{tr}[\hat{V}_\mu]\mathrm{tr}[T\hat{V}_\nu]X^{\mu\nu}
+ig\alpha_{40}\mathrm{tr}[\hat{V}^\mu]\mathrm{tr}[T\hat{V}^\nu W_{\mu\nu}] \end{eqnarray} among which
$\alpha_1,\alpha_8,\alpha_{24},\alpha_{25}$ are kinetic mixing parameters; $\alpha_{12}\sim\alpha_{14}$, $\alpha_{30}$,
$\alpha_{33}\sim\alpha_{40}$ are associated with CP-violation terms. The anomalous term $\mathcal{L}_A$ is \begin{eqnarray} \mathcal{L}_A
=\alpha_{42}g^2\epsilon^{\mu\nu\rho\lambda}tr[W_{\mu\nu}W_{\rho\lambda}]
    +\alpha_{43}g'^2\epsilon^{\mu\nu\rho\lambda}B_{\mu\nu}B_{\rho\lambda}
    +{g^{\prime\prime}}^2\alpha_{44}\epsilon^{\mu\nu\rho\lambda}X_{\mu\nu}X_{\rho\lambda}~~~~~
\end{eqnarray} With the exception of the kinetic term $\mathcal{L}_K$ in $\mathcal{L}_4$, the $\alpha_i$ ($i=1,\cdots,14$) correspond to terms
appearing in the conventional electroweak chiral Lagrangian (EWCL) \cite{CLold} without Z', $\alpha_j$ ($j=15,\cdots,44$) correspond to terms in
the EEWCL involving Z'. In the rest of the paper, we shall ignore the CP-violation terms and the anomaly terms. This constitutes our third and
fourth assumptions. Our fifth assumption is to forbid possible mixing between the Z' and the electroweak bosons $\gamma,Z$ in order to keep the
Z' dark. This implies that the gauge eigenstate $X_\mu$ can be identified with the physical state $Z'_\mu$. No mass mixing requires $\beta_2=0$,
and no kinetic mixing leads to $\alpha_{24}=\alpha_{25}=0$. Furthermore, we need to set to zero the Stueckelberg coupling $\tilde{g}' = 0$. At
this point it is important to stress that there are mixing parameters which do not involve the Z': $\beta_1$ is a mass mixing term among SM
gauge bosons, thus it will induce a correction to the $\rho$ parameter ($T$ parameter), while $\alpha_1$ and $\alpha_8$ induce kinetic mixings,
thus generating a contribution to the $S$ parameter. The remaining $\beta_3$ generates a contribution to the mass of the Z', which is given by
\begin{equation} M_{\rm Z'} = g'' f \sqrt{1-2 \beta_3}. \end{equation}

With the above five assumptions, the $p^4$-order EEWCL (\ref{Ldef}) in the unitary gauge gives the following Lagrangian up to quartic couplings:
\begin{eqnarray} \mathcal{L} &=&-\frac{1}{4}V_{\mu\nu}V^{\mu\nu} -\frac{1}{2}W^+_{\mu\nu}W^{-\mu\nu}
        +iC_{V-+}V_{\mu\nu}W^{+\mu}W^{-\nu}
    +iC_{+V-}(W^+_{\mu\nu}W^{-\mu}V^\nu-W^-_{\mu\nu}W^{+\mu}V^\nu)
    \nonumber\\
    &&+iC_{V_1V_2V_3}V^{\mu\nu}_1V_{2\mu} V_{3\nu}
    +D_{++--}W^+_\mu W^{+\mu}W^-_\nu W^{-\nu}
    +D_{+-+-}W^+_\mu W^{-\mu}W^-_\nu W^{+\nu}
    \nonumber\\
    &&+D_{+-V_1V_2}W^+_\mu W^{-\mu}V_{1\nu} {V_2}^\nu
    +D_{+V_1-V_2}W^+_\mu V_1^\mu W^-_\nu V_2^\nu
    +D_{V_1V_2V_3V_4}V_{1\mu}V_2^\mu V_{3\nu}V_4^\nu.
\end{eqnarray} Here $V_i$ denotes the neutral gauge bosons $Z$, $\gamma$, and $Z'$, and the various $C$ and $D$ coefficients in terms of the
$\alpha_i$ and $\beta_i$ coefficients are given in Appendix \ref{couplings}. Note that, with our fifth assumption of no mass mixing,
$C_{V_1V_2V_3}$ vanishes.

In order to keep the Z' stable, we need to impose the vanishing of vertices that are linear in the Z' field; this constitutes our sixth
assumption. For the triple couplings $C_{Z'-+}$ and $C_{+Z'-}$ to vanish, we need to set $\alpha_{31}=0$; there is then no triple coupling
involving $Z'$ (note that without the triple coupling $C_{Z'VV}$ with $V=\gamma,Z,W^{\pm}$, the longitudinal $W$ and $Z$ scattering will not
involve $Z'$ at tree level and that imposes no unitarity constraint on the Z' couplings). Left with four Z'-independent triple couplings, one
$C_{+\gamma-}$ has fixed coefficients, which only depend on SM couplings, whereas the other three $C_{\gamma-+},C_{Z-+}$, and $C_{+Z-}$ are
free, corresponding to independent coefficients $\alpha_2,\alpha_3$, and $\alpha_9$. For quartic couplings, $D_{+-ZZ'}=0$ leads to
$\alpha_{16}=0$, $D_{+Z-Z'}=D_{+Z'-Z}=0$ leads to $\alpha_{17}=0$, and $D_{Z'ZZZ}=-g_Z^3g''(2\alpha_{15}+\alpha_{16}+\alpha_{17})=0$ further
leads to $\alpha_{15}=0$. (Note that if the Stueckelberg coupling $\tilde{g}'$ does not vanish, it will also generate nonzero $D_{+-ZZ'}$,
$D_{+Z-Z'}$, $D_{+-AZ'}$, $D_{+A-Z'}$.) We are finally left with 16 nonzero quartic couplings. Among them, $D_{+-\gamma\gamma}$ and
$D_{+\gamma-\gamma}$ also have fixed coefficients and are not free. The other four $D_{+-\gamma Z}$, $D_{+\gamma -Z}$ and
$D_{+Z-\gamma}=D_{+\gamma-Z}$ only rely on $\alpha_3$ and then are related to the triple vertex. The remaining 11 nonzero quartic couplings
$D_{++--}, D_{+-+-}, D_{+-ZZ}, D_{+-Z'Z'}, D_{+Z-Z}, D_{+Z'-Z'}$, $D_{ZZZZ}, D_{Z'Z'ZZ}, D_{Z'ZZ'Z}, D_{Z'Z'Z'Z}, D_{Z'Z'Z'Z'}$ are free,
corresponding to the 11 independent coefficients $\alpha_4$ to $\alpha_7$, $\alpha_{10}$, and $\alpha_{18}$ to $\alpha_{23}$. In Table I, we
list details of all the triple and quartic couplings.

\begin{table}[htdp]\label{TripleQuarticCouplings1} \caption{List of triple and quartic couplings.} \begin{center} \begin{tabular}{c|c|c|c|c}
\hline\hline couplings & exist in SM (modified by)& independent & control by & vanishing condition\\ \hline $C_{\gamma-+}$ & yes
($\alpha_{2,3,9})$ & - & - & - \\ $C_{Z-+}$ & yes ($\alpha_{2,3,9}$) & - & - & - \\ $C_{Z'-+}$ & no & yes & $\alpha_{31}~\&~\tilde{g}'$ &
$\alpha_{31}=\tilde{g}'=0$ \\ $C_{+\gamma-}$ & yes (not modified)  & - & - & - \\ $C_{+Z-}$ & yes ($\alpha_3$) & - & - & - \\ $C_{+Z'-}$ & no &
- & $\tilde{g}'$ & $\tilde{g}'=0$ \\ $C_{V_1V_2V_3}$ & no & - & - & always \\ \hline $D_{++--}$ & yes ($\alpha_{3,4,9}$) & - & - & - \\
$D_{+-+-}$ & yes ($\alpha_{3,4,5,9}$) & - & - & - \\ $D_{+-\gamma\gamma}$ & yes (not modified) & - & - & - \\ $D_{+-ZZ}$ & yes
($\alpha_{3,5,7}$) & - & - & - \\ $D_{+-Z'Z'}$ & no & yes & $\alpha_{21}$ & - \\ $D_{+-\gamma Z}$ & yes ($\alpha_3$) & - & - & - \\ $D_{+-\gamma
Z'}$ & no & - & $\tilde{g}'$ & $\tilde{g}'=0$ \\ $D_{+-ZZ'}$ & no & yes & $\alpha_{16}~\&~\tilde{g}'$ & $\alpha_{16}=\tilde{g}'=0$ \\
$D_{+\gamma-\gamma}$ & yes (not modified) & - & - & - \\ $D_{+Z-Z}$ & yes ($\alpha_{3,4,6}$) & - & - & - \\ $D_{+Z'-Z'}$ & no & yes &
$\alpha_{19}$ & - \\ $D_{+\gamma-Z}$ & yes ($\alpha_3$) & - & - & - \\ $D_{+\gamma-Z'}=D_{+Z'-\gamma}$ & no & - & $\tilde{g}'$ & $\tilde{g}'=0$
\\ $D_{+Z-\gamma}=D_{+\gamma-Z}$ & yes & - & - & - \\ $D_{+Z-Z'}=D_{+Z'-Z}$ & no & yes & $\alpha_{17}~\&~\tilde{g}'$ &
$\alpha_{17}=\tilde{g}'=0$ \\ $D_{ZZZZ}$ & no & yes & $\alpha_{10}$ & - \\ $D_{Z'ZZZ}$ & no & yes & $\alpha_{15}$ &
$2\alpha_{15}+\alpha_{16}+\alpha_{17}=0$ \\ $D_{Z'Z'ZZ}$ & no & yes & $\alpha_{20}$ & - \\ $D_{Z'ZZ'Z}$ & no & yes & $\alpha_{18}$ & - \\
$D_{Z'Z'Z'Z}$ & no & yes & $\alpha_{22}$ & $\alpha_{22}=0$ \\ $D_{Z'Z'Z'Z'}$ & no & yes & $\alpha_{23}$ & - \\ \hline\hline \end{tabular}
\end{center} \label{default} \end{table}%

Given the six assumptions stated above: \begin{itemize} \item[(i)] no direct coupling to the Higgs; \item[(ii)] no direct coupling to fermions;
\item[(iii)] no CP violating terms in the EEWCL; \item[(iv)] no anomalous terms in the EEWCL; \item[(v)] no kinetic nor mass mixing terms;
\item[(vi)] no single Z' couplings; \end{itemize} we are finally left with four $Z'$-dependent quartic vertices each involving two $Z'$ fields,
\begin{eqnarray} D_{+-Z'Z'}&\equiv&g_1=4g^2g^{''2}(\alpha_5+\alpha_{21})\\ D_{+Z'-Z'}&\equiv&g_2=4g^2g^{''2}(\alpha_4+\alpha_{19})\\
D_{Z'Z'ZZ}&\equiv&g_3=g_Z^2g^{''2}(\alpha_5+2\alpha_7+4\alpha_{20}+2\alpha_{21})\\
D_{Z'ZZ'Z}&\equiv&g_4=4g_Z^2g^{''2}(\alpha_4+\alpha_6+2\alpha_{18}+\alpha_{19}) \end{eqnarray} where $g_Z^2 = g^2 + {g'}^2$. Here we limit
ourselves to vertices with up to 4 particles: the EEWCL does contain Z' vertices with more gauge boson, however their physical effect is
subleading and we will not consider them any further. The above couplings do contribute to electroweak precision tests at loop level: such
contributions are log--divergent, and their contribution can be absorbed in the tree--level contributions $\beta_1$ and $\alpha_{1,8}$, whose
values are therefore strongly constrained. In the following we will therefore not consider bounds from precision tests, as they do not give
unique indication on the size of the quartic couplings listed above. To keep the number of free parameters to a minimum, we shall take in our
analysis a unique coupling constant $g_0$ and consider five different arrangements of coupling constants as follows: \begin{itemize} \item
universal case:~$g_1\!=\!g_2\!=\!4g^2g_0$,~$g_3\!=\!\frac{3}{8}g_4\!=\!\frac{3}{2}g_Z^2g_0$,~or $\alpha_4\!=\!\alpha_5\!=\!
\alpha_{19}\!=\!\alpha_{21}\!\equiv\!\frac{g_0}{2g^{''2}}$,~$\alpha_6\!=\!\alpha_7\!=\!\alpha_{18}\!=\!\alpha_{19}\!=\!\alpha_{20}\!=0$ \item
case 1:~$g_1=4g^2g_0$,~$g_3=\frac{3}{2}g_Z^2g_0$,~ $g_2=g_4=0$,~or $\alpha_5=\alpha_{21}\equiv\frac{g_0}{2g^{''2}}$,~$
\alpha_4=\alpha_6=\alpha_7=\alpha_{18}=\alpha_{19}=\alpha_{20}=0$ \item case 2:~$g_2=4g^2g_0$,~$g_4=4g_Z^2g_0$,~$g_1=g_3=0$,~or
$\alpha_4=\alpha_{19}\equiv\frac{g_0}{2g^{''2}}$,~$\alpha_5
    =\alpha_6=\alpha_7=\alpha_{18}=\alpha_{20}=\alpha_{21}=0$
\item case 3:~$g_3=3g_Z^2g_0$,~$g_1=g_2=g_4=0$,~or $\alpha_7=\alpha_{20}\equiv\frac{g_0}{2g^{''2}}$,~$\alpha_4
    =\alpha_5=\alpha_6=\alpha_{18}=\alpha_{19}=\alpha_{21}=0$
\item case 4:~$g_4=6g_Z^2g_0$,~$g_1=g_2=g_3=0$,~or $\alpha_6=\alpha_{18}\equiv\frac{g_0}{2g^{''2}}$,~$\alpha_4
    =\alpha_5=\alpha_7=\alpha_{19}=\alpha_{20}=\alpha_{21}=0$.
\end{itemize} In the following sections, we shall mainly focus on the above five cases and derive information on the unique coupling $g_0$ when
discussing possible constraints from various experiments. Some possible exceptions are also discussed.

\section{Relic density constraint of dark Z'}

In the standard Cosmology picture ($\Lambda$CDM), it is assumed that the DM particles are in thermal equilibrium with the other SM particles via
various fundamental processes such as $Z'Z'\rightarrow\bar{P}P$ where $P$ is any SM particle. In the high-temperature Early Universe, DM
particles were kept in thermal equilibrium as long as the reaction rate, scaled by the temperature, was faster than the expansion rate $H$ (the
Hubble parameter) of the Universe. The Universe cooled down as it continued to expand. At temperatures around which the reaction rate fell below
the expansion rate $H$, the DM particles began to decouple from the thermal bath. The DM particles will continue to annihilate into SM particles
up until the point when they no longer encounter one another. The remaining DM particles will then become the relics that we can observe today.
The two possible annihilation processes for our dark Z' are shown in Fig.~1. \begin{figure} \begin{center}
\scalebox{0.2}{\includegraphics{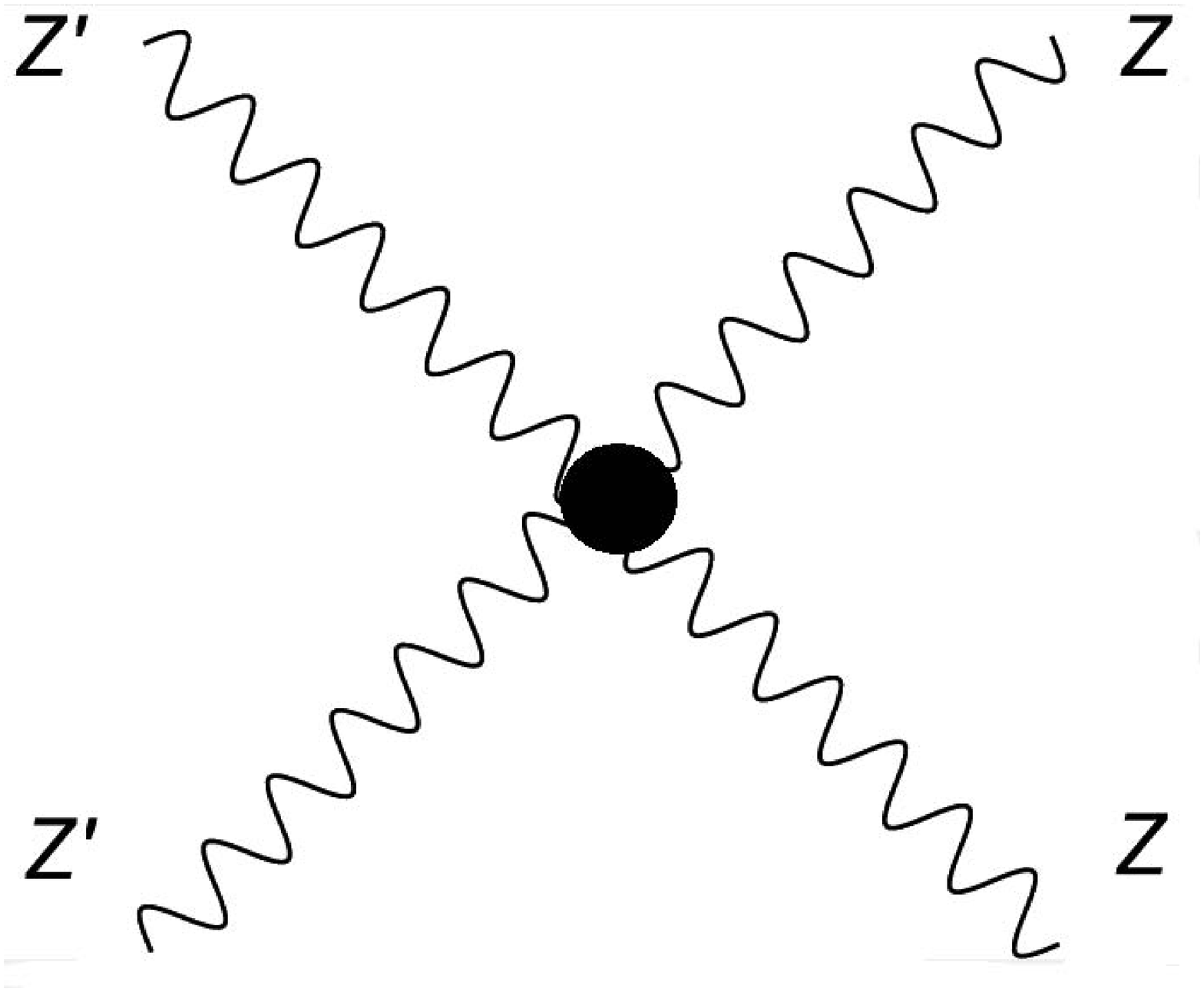}}\qquad \qquad \scalebox{0.2}{\includegraphics{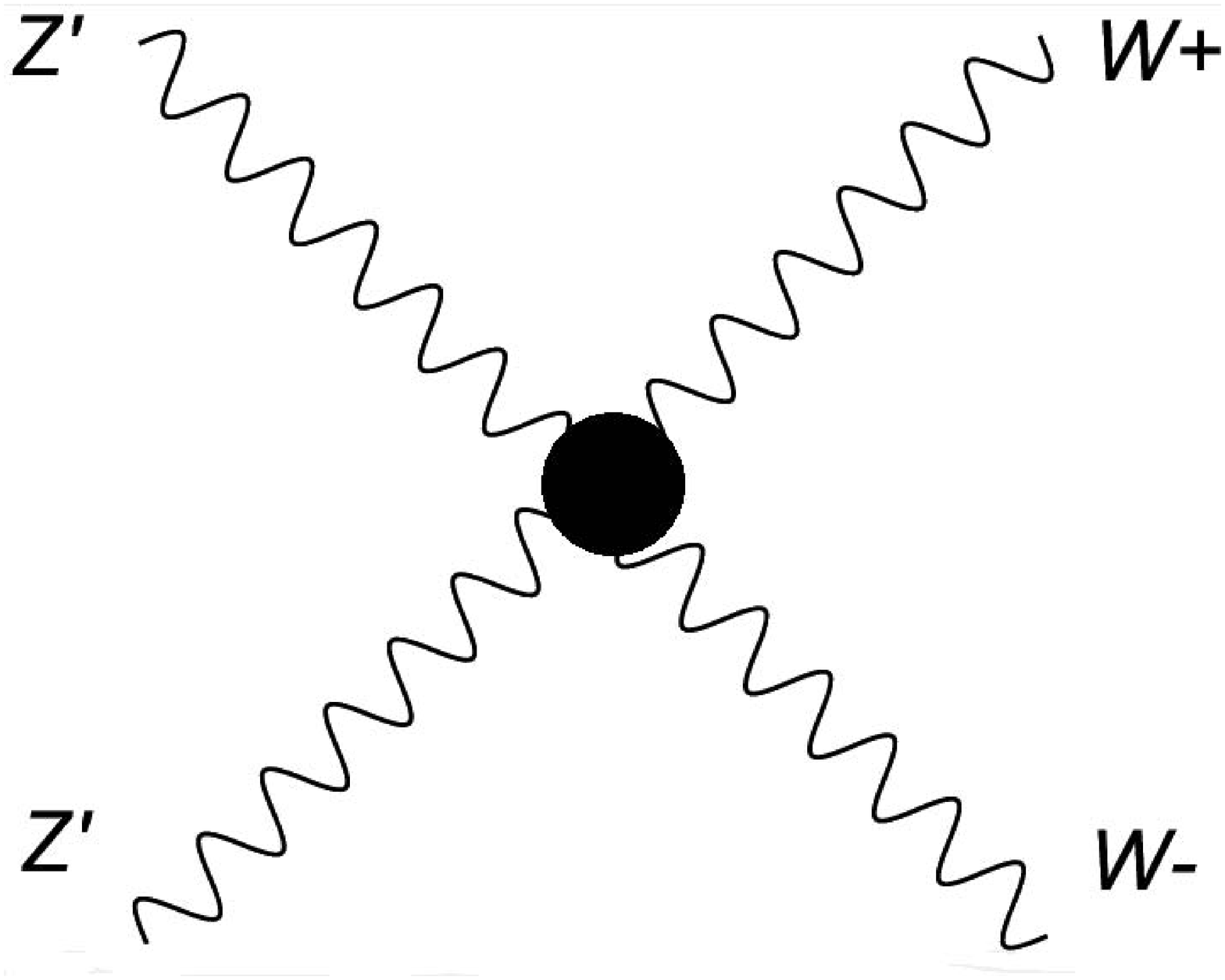}} \end{center} \caption{Two annihilation
processes of $Z'$ to SM weak gauge bosons.} \end{figure} Their annihilation rates to SM weak gauge bosons are \begin{eqnarray} \sigma_{W}\;
v&=&\frac{\sqrt{1-\frac{1}{r_W^2}}}{9\times 64\pi m_{Z'}}\{[(224r_W^4+112r_W^2+136)g_1^2-
(160r_W^4+80r_W^2+176)g_1g_2+(152r_W^4+128r_W^2+96)g_2^2]\nonumber\\
&&+[(88r_W^4-40r_W^2+56)g_1^2-(\frac{304}{3}r_W^4+32r_W^2+\frac{32}{3})g_1g_2+(24r_W^4+\frac{112}{3}r_W^2
-\frac{4}{3})g_2^2]v^2\}\label{SigmaWv}\\ \sigma_{Z}\; v&=&\frac{\sqrt{1-\frac{1}{r_Z^2}}}{18\times 64\pi
m_{Z'}}\{[(224r_Z^4+112r_Z^2+136)g_3^2- (160r_Z^4+80r_Z^2+176)g_3g_4+(152r_Z^4+128r_Z^2+96)g_4^2]\nonumber\\
&&+[(88r_Z^4-40r_Z^2+56)g_3^2-(\frac{304}{3}r_Z^4+32r_Z^2+\frac{32}{3}g_3g_4+(24r_Z^4+\frac{112}{3}r_Z^2
-\frac{4}{3})g_4^2]v^2\},\label{SigmaZv}
\end{eqnarray} in which $r_W=m_{Z'}/m_W$ and $r_Z=m_{Z'}/m_Z$, and $v$ is the relative velocity of the
colliding DM particles. In above result, due to the non-relativistic characteristics of our dark Z', we have taken expansion in terms of powers of $v$ up to the order of $v^2$.
Note that for these annihilations to arise, the mass of Z' should be heavier than the mass of the SM weak gauge bosons.
If the Z' mass is lighter than the SM weak boson mass, we instead use the loop-induced effective $Z'Z'q\bar{q}$ vertices given in the next
section. The corresponding annihilation rate is \begin{eqnarray} \sigma \; v&=&\frac{v^2}{6}\bigg[\frac{1}{6\pi}{\displaystyle\sum_f}
(\frac{K_{V,f}}{\sqrt{2}})^2c_f \sqrt{1-\frac{m_f^2}{M^2_{Z'}}}M^2_{Z'}(2+\frac{m_f^2}{M^2_{Z'}})+
\frac{1}{3\pi}{\displaystyle\sum_f}(\frac{K_{VA,f}}{\sqrt{2}})^2c_f(1-\frac{m_f^2}{M^2_{Z'}})^{3/2}M^2_{Z'}\bigg],\label{sigmav}
 \end{eqnarray}
where $K_{V,f}$ and $K_{VA,f}$ are the effective couplings introduced in the next section in Eq.~(\ref{VandAxialV}), and $c_f$ is the color
index, which is 1 for leptons and 3 for quarks.

The relic density is calculated by solving the Boltzmann equation in the standard approximation procedure \cite{Kolb:1981hk}, \begin{equation}
\Omega_{\mathrm{WIMP}}h^2=\frac{1.07\times10^9}{m_{pl}} \frac{x_FGeV^{-1}}{\sqrt{g_{*S}}}\frac{1}{a+3b/x_F}, \end{equation} where $h$ is the
scaled Hubble constant, $x_F=m_{Z'}/T_F$ with $T_F$ the freezing temperature, $m_{pl}=1.22\times 10^{19}GeV$, $g_{*S}$ the total number of
effectively relativistic degrees of freedom at freeze-out temperature, and $a$ and $b$ are parameters in the expansion $\sigma
v=a+bv^2+\mathcal{O}(v^4)$. The freeze-out temperature parameter $x_F$ can be evaluated by numerically solving the following equation:
\begin{equation} x_F=\ln \left[c(c+2)\sqrt{\frac{45}{8}\frac{gm_{Z'}m_{pl}(a+6b/x_F)}{2\pi^3\sqrt{g_{*S}}x_F^{1/2}}}\right], \end{equation}
where $g=3$ is the number of degrees of freedom for the $Z'$ DM, and $c$ is a numerical constant usually taken equal to $1/2$. With DM mass
ranging from GeV to TeV, $x_F\approx 25$ and remains essentially constant. In our numerical analysis, we demand that the resulting relic density
be less than the measured value from PLANCK $\Omega_{\mathrm{WIMP}}h^2=0.1199\pm0.0027$ at $68\%$CL \cite{PLANCK}, which leads to constraints
for the effective coupling constants $g_i,\ i=1,2,3,4$ and the dark Z' mass $M_{Z'}$. The result of the five different coupling arrangements
introduced at the end of Sec. II for $M_Z'>100$GeV is shown in Fig.~2, where we have used result (\ref{SigmaWv}) and (\ref{SigmaZv}) to perform
our calculation. Note that because the ordinate is logarithmically scaled, the diagram would be unable to show clearly the possible deviations
of several percent from the experiment, and hence are not plotted.
 \begin{figure}
\begin{center} \scalebox{0.4}{\includegraphics{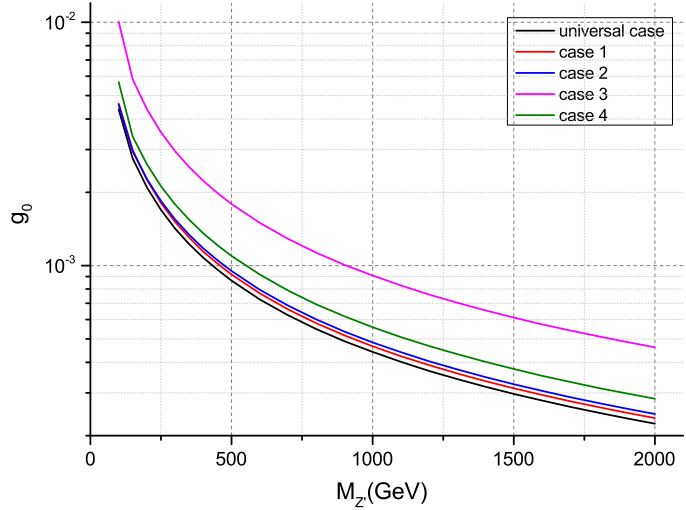}} \end{center} \vspace*{-0.5cm}\caption{Predicted unique coupling constant $g_0$ as a
function of DM mass $m_{Z'}\geq 100$GeV fixed by the observed relic density. Each color of the curves represents different arrangements of
coupling constant $g_i$. The allowed region is located above the curve.} \end{figure} For dark $Z'$ masses below the W mass threshold, we
instead use Eq.~(\ref{sigmav}) to perform our estimation; the results are shown in Fig.~3. \begin{figure} \begin{center}
\scalebox{0.4}{\includegraphics{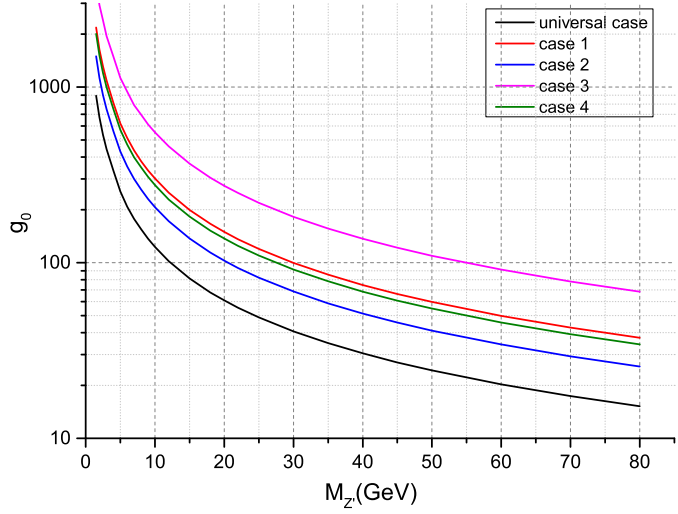}} \end{center} \vspace*{-0.5cm}\caption{Same as for Fig. 1 but with DM mass $m_{Z'}\leq 80$~GeV.}
\end{figure} Note that in the intermediate region $m_W/2 < M_{Z'} < m_W$, the three--body process $Z' Z' \to W W^\ast (Z Z^\ast )$, where one of
the two gauge bosons is off--shell, is also relevant: we expect it to smoothly interpolate between the results in Fig.~2 and 3. Nevertheless,
below the $WW$ threshold, the coupling is required to be very large, and this region is excluded by direct--detection experiments, as we will
show in the next section.

\section{Effective quark vertex and direct detection of dark Z'}

In this section, we discuss the direct-detection constraints on dark Z'. Direct-detection experiments are designed to measure the recoil energy
of the atomic nuclei following DM elastic scattering. DM-quark interactions will naturally induce DM-nucleon interactions, and the latter
further induce DM-nucleus interactions. Such interactions may be detected in underground direct-detection experiments.

As our second assumption does not allow $Z'$ to directly couple to SM fermions (fermiophobic), it can only couple indirectly to quarks through
the SM gauge-boson loop depicted in Fig.~4. \begin{figure} \begin{center} \scalebox{0.3}{\rotatebox{-90}{\includegraphics{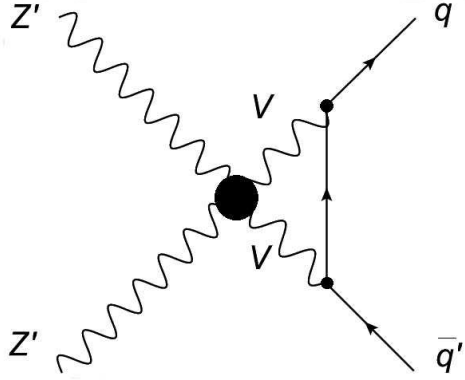}}}
\end{center} \caption{Dark matter scattering from quarks through SM gauge-boson loops.} \end{figure} This loop diagram leads to finite effective
vector-vector and vector-axial-vector interaction vertices \cite{Yu:2011by}: \begin{equation}
\mathcal{L}_{eff}=\Sigma_q\frac{K_{V,q}}{\sqrt{2}}(Z'^*_{\nu}i\overleftrightarrow{\partial}_{\mu}Z'^{\nu})\bar{q}\gamma^{\mu}q+
\Sigma_q\frac{K_{VA,q}}{\sqrt{2}}(Z'^*_{\nu}i\overleftrightarrow{\partial}_{\mu}Z'^{\nu})\bar{q}\gamma^{\mu}\gamma_5q.\label{VandAxialV}
\end{equation} The finiteness of this loop-induced $Z'Z'\bar{q}q$ vertex implies that our fermiophobic assumption is consistent by itself.
Furthermore, the structure of the above result for the effective vertices only generates spin-independent (SI) interaction. Hence there will be
no constraint from the cross-section of the spin-dependent interaction.

Note that the velocity of DM near the Earth is considered to be $v\approx 0.001c$; for low-energy SI interactions, only the vector interaction
survives. The cross-section is then given by \begin{equation} \sigma_{V,Z'N}=\frac{m_N^2m_{Z'}^2}{\pi(m_{Z'}+m_N)^2}
(\frac{K_{V,N}}{\sqrt{2}})^2,\label{CrossSectionVector} \end{equation} where \begin{eqnarray} &&K_{V,p}=2K_{V,u}+K_{V,d},\quad
K_{V,n}=K_{V,u}+2K_{V,d}\label{Kvp}\\
&&\frac{K_{V,q}}{\sqrt{2}}=\frac{(g_1+g_2)(c_q^2+c_q'^2)}{32\pi^2m_{W}^2}+\frac{(g_3+g_4)(c_q^2+c_q'^2)}{32\pi^2 m_{Z}^2}\label{Kvq}
\end{eqnarray} and $c_q$ and $c_q'$ for $q=u,d$ are coefficients associated with SM couplings \begin{equation}
c_u=\frac{ig}{4\cos\theta_W}(1-\frac{8}{3}\sin^2\theta_W),\quad c_d=\frac{ig}{4\cos\theta_W}(-1+\frac{4}{3}\sin^2\theta_W),\quad
c_u'=\frac{ig}{4\cos\theta_W},\quad c_d'=-\frac{ig}{4\cos\theta_W}. \end{equation}

The curves for the predicted cross-section $\sigma$ vs $m_{Z'}$ with different coefficient settings are shown in Fig.~5. The upper bounds set by the XENON100 2012 data \cite{XENON100}, the latest LUX result \cite{LUX}  and SuperCDMS result \cite{SuperCDMS} are also plotted in Fig.~5. \begin{figure} \begin{center}
\scalebox{0.4}{\includegraphics{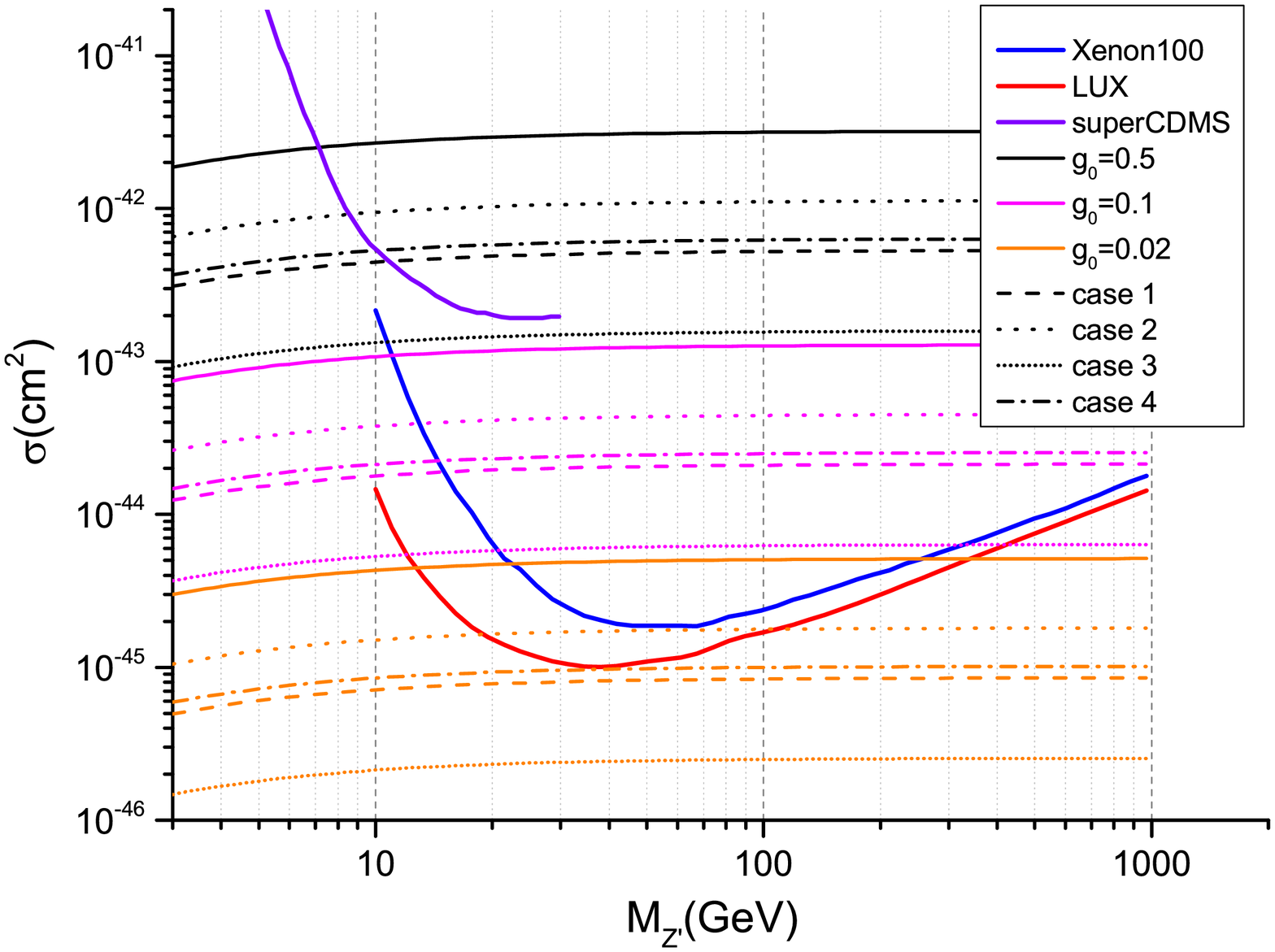}} \end{center} \caption{Predicted spin-independent WIMP-proton cross-sections for dark Z'. The
solid line is for the universal case, dashed line for case 1, dotted line for case 2, short dotted line for case 3, and dash-dotted line for
case 4. For each case, the black lines correspond to $g_0=0.5$, magenta lines to $g_0=0.1$, and orange lines to $g_0=0.02$. Also, the upper
bounds set by XENON100, LUX and SuperCDMS are marked in blue, red and purple respectively.} \end{figure}

We next discuss some of these results: \begin{itemize} \item[-] Because the unique coupling $g_0$ is constrained to order $10^{-2}$ in the typical few tens to few hundreds GeV region (see
Fig.~5), it is natural to ask if we go beyond the limitations set for the five arrangements and consider arbitrary $g_i$ couplings; is there a
possibility to enhance the value of the coupling? To examine this issue, note that Eq.~(\ref{Kvq}) gives $K_{V,q}$, which then determines the
final cross-section $\sigma$; it depends on $g_1+g_2$ and $g_3+g_4$. If we take $g_1\approx-g_2=4g^2g_0$, $g_3\approx-g_4=4g_Z^2g_0$, then we
achieve an arbitrary small $K_{V,q}$ and subsequently an arbitrary small scattering cross-section. For example, $g_2=-0.9g_1$, $g_4=-0.9g_3$,
and $g_1=g_3$ leads to $K_{V,q}\propto 0.1g_1$, which is already smaller by an order of magnitude than the five cases already derived in Fig.~5.
Hence $\sigma\propto 0.01g_1$ is less than those used in Fig.~5 by two orders of magnitudes and is equivalent to enhancing the constraint for
$g_0$ by two order of magnitudes; i.e., it relaxes the constraint for $g_0$ from $10^{-2}$ to $10^0$, because experimentally we must fix the
cross-section. Therefore, we do have flexibility in the couplings in relaxing the constraints on $g_0$ (or $\alpha_i$) from direct detection. In
the next section, we shall see for indirect detection that this scenario does not arise as there exists no such coupling space.

\item[-] As we have introduced in our theory the one-loop-induced $Z'Z'\bar{q}q$ vertex (\ref{VandAxialV}), one may inquire of the role of the
    one-loop-induced $Z'Z'h$ vertex depicted in Fig.~6. One can easily check that this loop diagram is logarithmically divergent. This implies
    that our first higgsphobic assumption, introduced in Sec.II, by which we ignore the direct coupling between Z' and Higgs is not consistence
    by itself. To cancel such a divergence, we have to introduce into the theory the tree-level vertex, $Z'Z'h$, which violates our higgsphobic
    assumption. \begin{figure} \begin{center} \scalebox{0.4}{\rotatebox{0}{\includegraphics{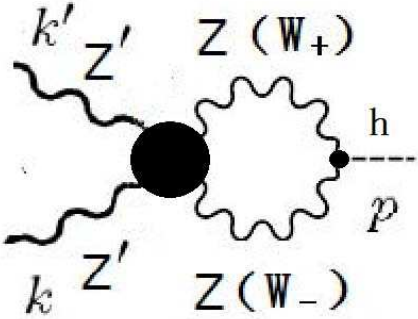}}} \end{center} \caption{Dark matter
    annihilating into Higgs through SM gauge boson loops.} \end{figure} Furthermore, once we have such a $Z'Z'h$ vertex, no matter whether it is
    a tree-level one added to the theory by hand or a loop-induced one, it can further decay via the Higgs-mediated s-channel (Fig.~7) into a
    fermion pair, which then leads to further corrections to the effective $Z'Z'\bar{q}q$ vertex given by Eq.~(\ref{VandAxialV}). \begin{figure}
    \begin{center} \scalebox{0.4}{\rotatebox{0}{\includegraphics{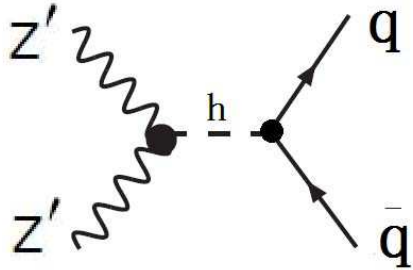}}} \end{center} \caption{Dark matter annihilating to form a quark
    pair through the effective $Z'Z'h$ vertex.} \end{figure} The reason that we do not consider this correction is as follows: suppose we
    abandon our first assumption by introducing the tree-level $Z'Z'h$ vertex into our theory. The $Z'Z'h$ vertex then includes two
    contributions: one is a tree-level term, and the other a loop term (Fig.~6). After renormalisation, i.e., cancelling the loop-induced
    divergence (Fig.~6) by the tree-level term, the remaining finite part should not be very much different with respect to the finite part of
    Fig.~6. This finite part of the $Z'Z'h$ vertex via the Higgs-mediated s-channel (Fig.~7) further leads to an effective $Z'Z'\bar{q}q$
    vertex, which is of scalar type $Z^{'2}\bar{q}q$ or pseudo-scalar type $Z^{'2}\bar{q}\gamma_5q$. These vertices have different vertex
    structures in comparison with the vector and axial vector vertices given by Eq.~(\ref{VandAxialV}). Only the scalar vertex survives and it
    contributes to the nucleon cross-section similar to that in (\ref{CrossSectionVector}), but has an extra factor proportional to
    $m_q^2/m_{Z'}^2\sim 10^{-8}$ with $m_q$ the mass of the u or d quark. This factor results from the replacement of the derivative-type vector
    coupling of (\ref{CrossSectionVector}) with the non-derivative-type scalar coupling $Z'Z'\bar{q}q$ (see Ref.\cite{Yu:2011by}). It is this
    suppression factor that allows us to ignore the loop-induced and tree-level $Z'Z'h$ vertices. Although our higgsphobic assumption is not
    consistent by itself, ignoring it only creates a very minor correction and thus we can still approximately adhere to it. \end{itemize}

\section{Indirect detection of dark Z'}

In addition to the direct DM searches at underground laboratories, indirect searches look for DM annihilations or decay products in the
atmosphere. These particles, which include neutrinos, gamma rays, positrons, and antiprotons, can be detected in cosmic-ray experiments. In this
section, we discuss three of the latest experiments.
\subsection{PAMELA experiment}

We first follow the procedure in \cite{Yu:2011by} to obtain constraints for Z' using the anti-proton to proton flux ratio $\bar{p}/p$ measured
by the satellite-borne experiment PAMELA \cite{Adriani:2010rc}. The tree-level annihilations $Z'Z'\rightarrow ZZ,WW$ contribute to the
generation of the $\bar{p}/p$ signal via the hadronic decays of the W and Z bosons \cite{ppbar}. Together with the antiproton-to-proton flux
ratio data of PAMELA, we can derive constraints for each of our five cases. The result is shown in Fig.~8, where the allowed region is below
each curve. \begin{figure} \begin{center} \scalebox{0.4}{\includegraphics{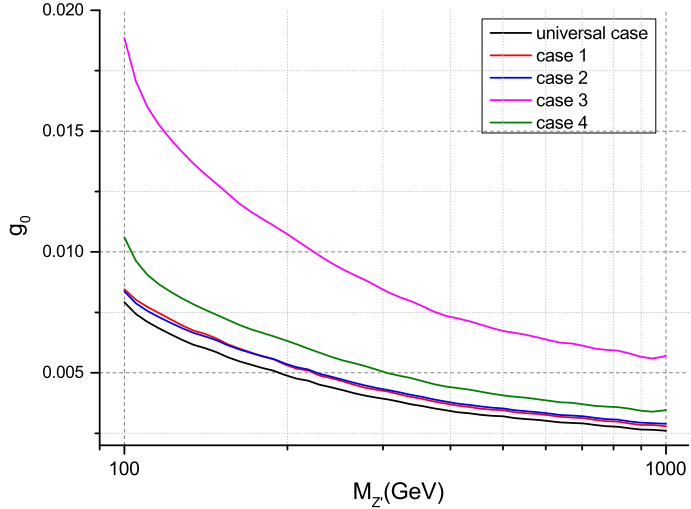}} \end{center} \vspace*{-0.5cm}\caption{Upper bounds on the
coupling constant $g_0$ from the PAMELA $\bar{p}/p$ flux ratio.} \end{figure} It should be noted that the loop-induced vertex Z'Z'qq given by
Eq.~(\ref{VandAxialV}) can also contribute to the generation of the $\bar{p}/p$ signal. Nevertheless, our computations show that the constraint
is very weak and we shall not discuss it here.
\subsection{AMSII experiment}

Recently, the AMSII collaboration has announced a new measurement of the cosmic-ray positron fraction \cite{AMS02}. We discuss DM annihilation
in view of these measurements and derive constraints on our dark Z' matter mass and the universal case for coupling constant $g_0$. We consider
the annihilation channels $Z'Z'\longrightarrow W^+W^-$ and $Z'Z'\longrightarrow ZZ$ as in the discussion for the relic density in the previous
section, and use the same method in \cite{AMS02explanation} to set conservative limits by requiring that the predicted positron flux remains
smaller than the measured flux over all energies. For simplicity, we assume a standard Navarro-Frenk-White (NFW) density profile
\cite{Navarro:1996gj}, i.e., $\bar{J}\Delta\Omega\approx 1$. The result is graphed in Fig.~9, which shown that it is not competitive with the
PAMELA bounds. \begin{figure} \begin{center} \scalebox{0.4}{\includegraphics{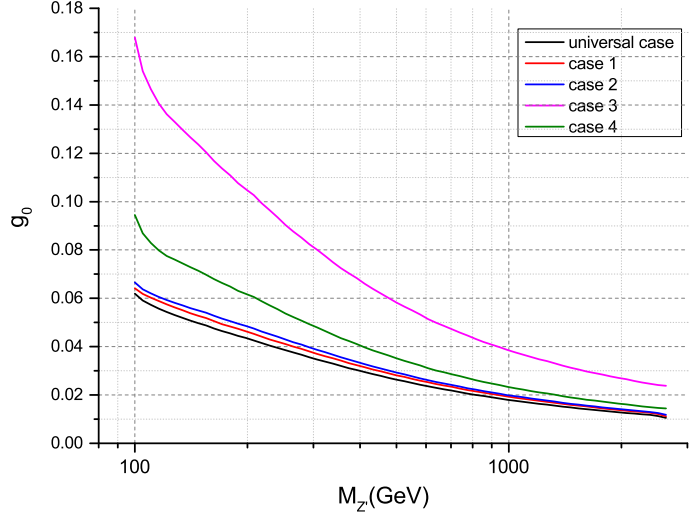}} \end{center} \vspace*{-0.5cm}\caption{Upper bounds from
AMSII cosmic-ray positron fraction spectrum.} \end{figure}

\subsection{FermiLAT experiment}

Photons from DM annihilations in the centre of the galaxy provide another source of an indirect signal. As dark Z' cannot annihilate into photon
pairs directly (we have checked that even by including one-loop corrections, the $Z'Z'\gamma$ vertex still vanishes), we can only detect
continuum photon signals. The differential flux of the $\gamma$-rays observed on Earth from DM annihilation is as follows, \begin{equation}
\frac{d\Phi}{dE_{\gamma}}=(5.5\times10^{-10}s^{-1}cm^{-2}) \frac{dN}{dE_{\gamma}}(\frac{\langle\sigma v\rangle}{pb})
(\frac{100GeV}{m_{\chi}})^2\bar{J}\Delta\Omega. \end{equation} Again, as in the previous subsection, we have assumed the DM distribution to
follow the NFW profile. The annihilation cross-section is \begin{equation} \langle\sigma v\rangle\approx a+b\langle v^2\rangle =a+2b\bar{v}^2,
\end{equation} where $\bar{v}=270km/s$. The simple analytic fit is as follows \cite{Bergstrom:1997fj} \begin{equation}
\frac{dN}{dE_{\gamma}}=\frac{dN}{m_{\chi}dx}=\frac{1}{m_{\chi}}\frac{a_0}{x^{1.5}}e^{-b_0x}, \end{equation} where $x=E_{\gamma}/m_{\chi}$,
$a_0=0.73$, and $b_0=7.76$ for $W/Z$ bosons. We consider the universal case with $g_0$ fixed by the relic density for different DM masses; the
predicted $\gamma$-ray spectra is shown in Fig.~ 10. \begin{figure} \begin{center} \scalebox{0.4}{\includegraphics{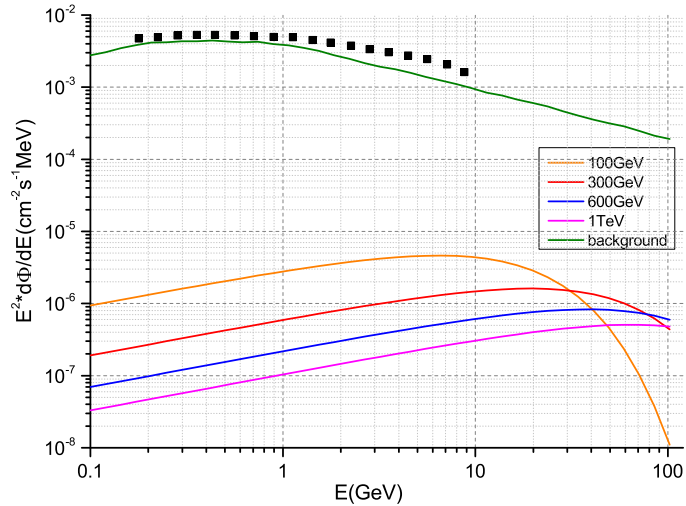}} \end{center}
\vspace*{-0.5cm}\caption{Predicted $\gamma$-ray spectra for the annihilation of dark Z' in the universal case and with a NFW density profile.
The FermiLAT observation data are also presented.} \end{figure} From it we can see the photon energy flux is about 3 orders of magnitude lower
than the experiment data. To explain the discrepancy with the data, we will need an enhancement of 300 to 8000. The required boost factors are
obtained by fitting the data and the result is shown in Fig.~11. \begin{figure} \begin{center} \scalebox{0.4}{\includegraphics{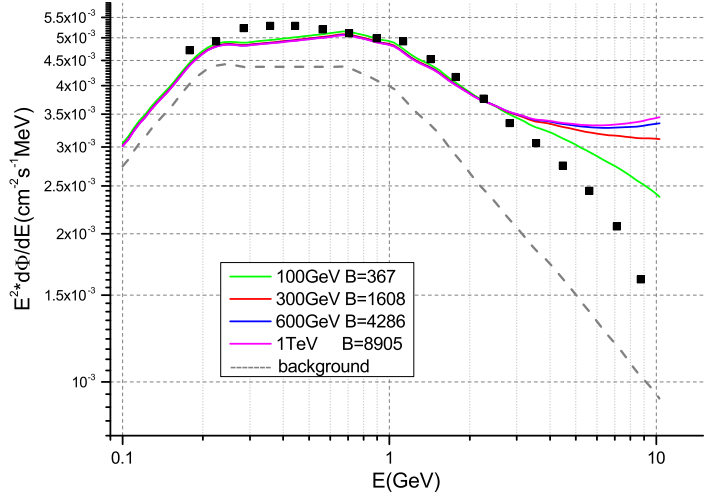}}
\end{center} \caption{Comparison of predicted dark Z' signals for different masses and boost factors with FermiLAT observation data.}
\end{figure}

Recently, $\gamma$-ray observations of 25 Milky-Way dwarf spheroidal satellite galaxies from four years of FermiLat data was reported in
\cite{Ackermann:2013yva}. We can use the constraint for the channel $Z'Z'\rightarrow W^+W^-$ in \cite{Ackermann:2013yva} to give upper limits to
coefficients $g_1$ and $g_2$. The result is shown below in Fig.~12. \begin{figure} \begin{center} \scalebox{0.4}{\includegraphics{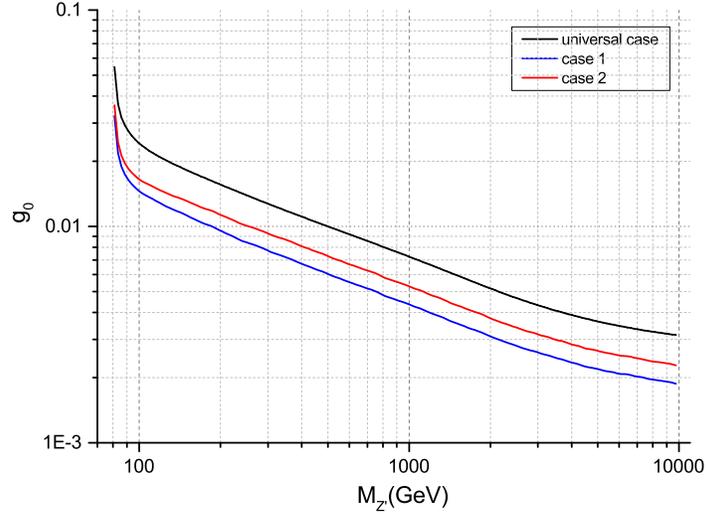}}
\end{center} \vspace*{-0.5cm}\caption{Upper limits to the dark-Z' coupling constant $g_0$ for the universal case, case 1, and case 2 from
FermiLAT $\gamma$-ray data.} \end{figure}

With the above discussion of the five $g_0$ cases, we may ask a similar question to that of the last section: if we go beyond the five limit
cases discussed above and consider arbitrary $g_i$ couplings, is there a possibility to enhance the value of the couplings? To answer this
question, note that the contributions to cross-section for indirect detection are given by Eq.~(\ref{SigmaWv}) and Eq.~(\ref{SigmaZv}), in which
the terms proportional to $v^2$ play very little role as $v^2$ is small. We can demand that the remaining terms take minimum values (which
result in the smallest indirect-detection cross-section) to fix the coupling. For $m_{Z'}=100~GeV$ (the result changes little in the range
100~GeV$<m_{Z'}<$1~TeV), we find two minima: \begin{eqnarray} g_2\approx 0.521g_1,~~~g_3=g_4=0 \hspace{1cm}\mbox{or} \hspace{1cm}
g_1=g_2=0,~~~g_4\approx 0.5207g_3. \end{eqnarray} This result gives the same sign for $g_1$ and $g_2$, and for $g_3$ and $g_4$. This differs
from the result in the direct detection discussed in the last section, where we obtain opposite signs for $g_1$ and $g_2$ and for $g_3$ and
$g_4$. Further, in plotting the constraint for these two extreme cases (Fig.~13), we find that the PAMELA experiment still yields constraint
$g_0<3\times10^{-2}$. This shows that, all in all, the constraint on the coupling is rather robust. \begin{figure} \begin{center}
\scalebox{0.4}{\includegraphics{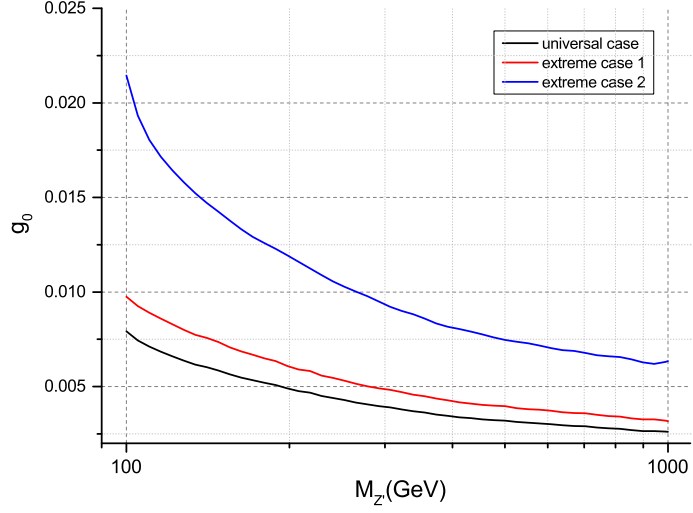}} \end{center} \vspace*{-0.5cm}\caption{Upper bounds from the PAMELA experiment on the coupling
constant $g_0$ for dark Z' in two extreme cases compared with the universal case.} \end{figure}

\section{Combined result and other DM related issues}

As the Z' is the only DM particle in our theory, we can discuss its mass $M_{Z'}$ and unique coupling $g_0$ in a $g_0-M_{Z'}$ plot. For
simplicity, we only discuss the universal case. Combining all effective phenomenological constraints from the last three sections, we obtain
Fig.~14, where the shaded region is excluded by the different experiments. \begin{figure} \begin{center}
\scalebox{0.4}{\includegraphics{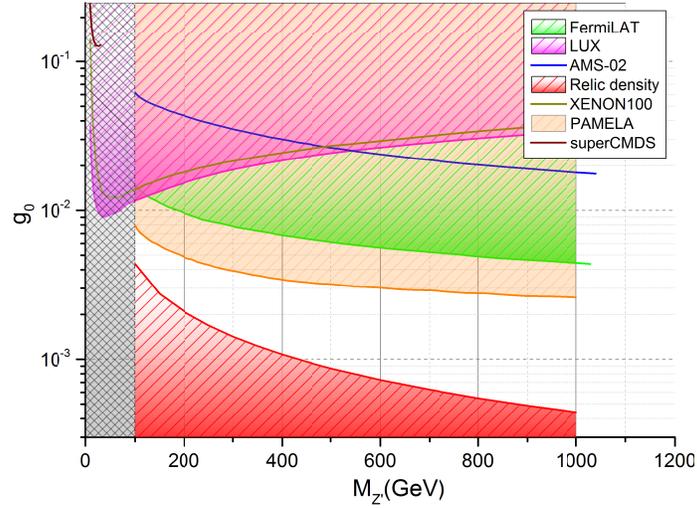}} \end{center} \vspace*{-0.5cm}\caption{Combined constraints on coupling constants $g_0$ of
dark Z' for the universal case.} \end{figure} The allowed region has a the left boundary at roughly $M_{Z'}\sim100$ GeV: below this mass scale,
the relic density constraint from Fig.~3 already contradicts data from the XENON100 and LUX experiments. The lower bound above $M_{Z'}=100$ GeV
is the red curve from the relic density (Fig.~2), while the upper bound is from the PAMELA $\bar{p}/p$ flux ratio constraint (Fig.~8). We find
$g_0$ is restricted to the region of $10^{-3}\sim 10^{-2}$. As discussed in the previous sections, this bound is quite solid, and it does depend
mildly on the configuration of couplings.

Considering the low-energy region below the threshold of the W boson mass, the relic density from Fig.~3 offers a very strong constraint, which
combined with the direct-detection results kills any possibility of the existence of a dark Z' in this low mass region. If we simply ignore this
fact, then a small dark Z' mass is allowed in this lower energy region. Currently, though, tensions exist among the different direct-detection
experiments in the low-energy region. For example, null results from CDMS Ge, XENON, LUX and SuperCDMS challenge the CoGeNT/DAMA and latest CDMS SI results. Because different experiments use different detection materials with different numbers of proton/neutrons, isospin-violating DM may
weaken these tensions. It was demonstrated that isospin-violating DM with $f_p/f_n\approx $-$0.7$ may alleviate the problem \cite{Feng:2011vu}.
For our dark Z', $f_q$ is given as follows: \begin{eqnarray}
&&\frac{f_q}{\sqrt{2}}=\frac{(g_1+g_2)(c_q^2+c_q'^2)}{32\pi^2m_W^2}+\frac{(g_3+g_4)(c_q^2+c_q'^2)}{32\pi^2m_Z^2}\\ &&f_p=2f_u+f_d,\hspace*{2cm}
f_n=f_u+2f_d. \end{eqnarray} One can check that \begin{equation} \frac{f_u}{f_d}=\frac{c_u^2+c_{u'}^2}{c_d^2+c_{d'}^2}, \end{equation} which
further leads to \begin{equation} f_p/f_n\approx 0.93. \end{equation} Clearly, the ratio is completely unrelated to the coupling constants
$g_1,g_2,g_3$, and $g_4$. Our dark Z' itself cannot give $f_p/f_n<0$ (adding in Higgs boson may bring new vertices and change the rate). Indeed,
it was recently observed that the LUX \cite{LUX} and SuperCDMS \cite{SuperCDMS} data are in strong contrast \cite{tension1,*tension2,*tension3} with the CDMS-Si signal result
\cite{CDMS-Si}, even for isospin- violating DM (IVDM) where the ratio of the proton-to-neutron couplings can be maximally ``xenophobic''.
Another possibility to reconcile this tension is to consider an inelastically scattering DM model or an exothermic DM model \cite{ExothermicDM}.
However, to realise inelastic scattering, we need at least two DM states. As our simplest dark Z' model only has one state, the present dark Z'
cannot produce inelastic scattering. This statement can be weakened in models where the Z' belongs to a triplet, i.e. it is accompanied by a W':
this can be achieved in a chiral lagrangian with an extra SU(2) symmetry~\cite{Wprime}. Remaining in our framework, we obtain the result that
even if we ignore the strong constraint allowing our dark Z' to survive into the low-energy region, it still cannot reduce the tension among
those observed possible DM signals with other null result experiments.

Finally, we need to discuss the impact of our dark Z' model on other experiments in particle physics. First, following our second and sixth
assumptions, there are no mixing of the Z' with the Z and photon and no linear couplings of the Z'. This implies that there are no tree level
bounds on the Z' couplings and mass from electroweak precision measurements\cite{EWprecision}. However, loops of the $Z'Z'WW$ and $Z'Z'ZZ$
couplings do generate corrections to both $T$ and $S$ parameters. Such loop corrections are logarithmical divergent, and they lead to the
renormalisation of the parameters $\beta_1$ and $\alpha_{1,8}$ in the EEWCL. Therefore, the new physics effects must be encoded in the
renormalised value of those parameters, and no robust and model independent bound on the couplings of the Z' can be extracted. Furthermore, the
allowed window for the coupling $g_0$ provides enough suppression to evade eventual bounds from the finite contribution of the loops. Another
issue regards the discovery potential at the LHC: as the only tree level couplings at the order we are interested in involve heavy gauge
bosons, the only channels where the Z' can be produced are production in association with a gauge boson ($p p \to Z' Z' (W,Z)$) and vector boson
fusion production ($p p \to Z' Z' j j$). One may therefore look for a single $W/Z$ signal or monojet. However, these channels suffer from large
backgrounds, and we checked that the small allowed values of the couplings ($g_0 \sim 10^{-3}$--$10^{-2}$) would lead to cross sections that
are too small to be detected at the LHC. We therefore conclude that the minimal model we study in this paper is not accessible at the LHC.

\section{Summary}

In this paper, we have investigated a rarely discussed possibility where a Z' boson is the sole DM candidate. \\We considered an extended chiral
Lagrangian with an additional U(1) gauge symmetry, with the following additional assumptions: \begin{enumerate} \item dark Z' is higgsphobic,
i.e., it does not directly couple to the Higgs, \item dark Z' is fermiophobic, i.e., it does not directly couple to SM fermions, \item there is
no CP violating Z' couplings, \item there is no anomalous Z' couplings, \item dark Z' does not mix with $\gamma$ and the $Z$ boson, and \item
there is no Z' interaction linear in the Z' field. \end{enumerate} The remaining quartic vertices, $Z'Z'ZZ$ and $Z'Z'W^+W^-$, then dominate the
Z' physics, which has four independent effective coupling constants $g_1,g_2,g_3,g_4$. We found that the mass of this dark Z' is not allowed
below the W boson mass threshold, due to a combination of strong constraints from the relic density and those from direct-detection experiments.
For mass $M_{Z'}>100$GeV, from the relic density and direct and indirect-detection experiments where effective $Z'Z'q\bar{q}$ couplings are
induced from SM gauge-boson loops, we produce five different coupling scenarios that are in the region $10^{-3}$--$10^{-2}$ (for the universal
case, the result is given in Fig.~14). This range of coupling can be relaxed beyond the five cases analyzed for direct-detection experiments,
but cannot be changed for indirect-detection experiments. To improve FermiLAT $\gamma$-ray spectra by our dark Z', we require a boost factor
from 300 to 8000. We checked that even if our dark Z' mass lies within the low-energy region, it cannot reduce tensions among the observed
possible DM signals with other null-result experiments. The bounds we extracted are therefore rather robust and model-independent.

 \section*{Acknowledgments}
This work was supported by the National Science Foundation of China (NSFC) under Grant No. 11075085, National Basic Research Program of China
(973 Program) under Grant No. 2010CB833000, the Specialized Research Fund for the Doctoral Program of High Education of China No.
20110002110010, France China Particle Physics Laboratory (FCPPL) financial support, and the Tsinghua University Initiative Scientific Research
Program. GC and AD also acknowledge partial support from the Labex-LIO (Lyon Institute of Origins) under grant ANR-10-LABX-66. AD is partially
supported by Institut Universitaire de France.

\appendix \section{List of Couplings}\label{couplings} \vspace*{-0.8cm}\begin{eqnarray} C_{\gamma-+}
&=&-\frac{g^3g'}{g_Z}(\frac{1}{g^2}+\alpha_2+\alpha_3+\alpha_9) \nonumber\\ C_{Z-+} &=&-\frac{g^2}{g_Z}[1-g'^2\alpha_2+g^2(\alpha_3+\alpha_9)]
\nonumber\\ C_{Z'-+} &=&-2{g}^{2}g''\alpha_{{31}} \nonumber\\ C_{+\gamma-} &=&-\frac{gg'}{g_Z} \nonumber\\ C_{+Z-} &=&-\frac{g^2}{g_Z}
    -g^2g_Z\alpha_3
\nonumber\\ C_{+Z'-} &=&0 \nonumber\\ C_{V_1V_2V_3}&=&0 \end{eqnarray} Here, $g_Z\equiv\sqrt{g_0^2+g_1^2}$. \begin{eqnarray} D_{++--}
&=&\frac{g^2}{2}+g^4(\alpha_3+\frac{\alpha_4}{2}+\alpha_9) \nonumber\\ D_{+-+-}
&=&-\frac{g^2}{2}+g^4(-\alpha_3+\frac{\alpha_4}{2}+\alpha_5-\alpha_9) \nonumber\\ D_{+-\gamma\gamma} &=&-\frac{g^2g'^2}{g_Z^2} \nonumber\\
D_{+-ZZ} &=&-\frac{g^4}{g_Z^2}
    -2g^4\alpha_3
    +g^2g_Z^2(\alpha_5+\alpha_7)
\nonumber\\ D_{+-Z'Z'} &=&4{g}^{2}{g''}^{2}(\alpha_{{5}}+\alpha_{21})\nonumber\\ D_{+-\gamma Z} &=&-\frac{2g^3g'}{g_Z^2}
    -2g^3g'\alpha_3
\nonumber\\ D_{+-\gamma Z'} &=&0 \nonumber\\ D_{+-ZZ'} &=&-2{g}^{2}g_Zg''\alpha_{{16}} \nonumber\\ D_{+\gamma-\gamma} &=&\frac{g^2g'^2}{g_Z^2}
\nonumber\\ D_{+Z-Z} &=&\frac{g^4}{g_Z^2}
    +2g^4\alpha_3
    +g^2g_Z^2(\alpha_4+\alpha_6)
\nonumber\\ D_{+Z'-Z'} &=&4{g}^{2}{g''}^{2}(\alpha_{{4}}+\alpha_{{19}}) \nonumber\\ D_{+\gamma-Z} &=&\frac{g^3g'}{g_Z^2}
    +g^3g'\alpha_3
\nonumber\\ D_{+\gamma-Z'} &=&D_{+Z'-\gamma} =0 \nonumber\\ D_{+Z-\gamma} &=&D_{+\gamma-Z}=\frac{g^3g'}{g_Z^2}+g^3g'\alpha_3 \nonumber\\
D_{+Z-Z'} &=&D_{+Z'-Z}=-{g}^{2}g_Zg''\alpha_{{17}} \nonumber\\ D_{ZZZZ} &=&\frac{1}{4}g_Z^4(\alpha_4+\alpha_5+2\alpha_6+2\alpha_7+2\alpha_{10})
\nonumber\\ D_{Z'ZZZ} &=&-g_Z^3g''(2\alpha_{15}+\alpha_{{16}}+\alpha_{17}) \nonumber\\ D_{Z'Z'ZZ}
&=&{g''}^{2}g_Z^2(\alpha_{{5}}+2\alpha_7+4\alpha_{20}+2\alpha_{21}) \nonumber\\ D_{Z'ZZ'Z}
&=&4{g''}^{2}g_Z^2(\alpha_{{4}}+\alpha_6+2\alpha_{18}+\alpha_{19}) \nonumber\\ D_{Z'Z'Z'Z}
&=&-4{g''}^{3}g_Z(\alpha_{{16}}+\alpha_{17}+2\alpha_{22}) \nonumber\\ D_{Z'Z'Z'Z'}
&=&4{g''}^{4}(\alpha_{{4}}+\alpha_5+2\alpha_{19}+2\alpha_{21}+4\alpha_{23}) \nonumber \end{eqnarray}

\end{document}